\documentclass{aa}  
\usepackage{graphicx}
\usepackage{txfonts}
\usepackage{longtable}
\begin{document}
   \title{A ``diamond-ring'' star: the unusual morphology of a young (multiple?) object\thanks{Based on observations collected at the ESO 8.2-m VLT-UT1 Antu telescope (programs 072.B-0335, 072.C-0335, 066.C-0015, and 063.I-0097).}}


   \author{Jo\~ao L. Yun\inst{1}
          \and
          Jos\'e M. Torrelles\inst{2}
	\and
	Nuno C. Santos\inst{1}
          }


   \institute{Universidade de Lisboa - Faculdade de Ci\^encias \\ Centro de Astronomia e Astrof\'{\i}sica da Universidade de Lisboa, \\
Observat\'orio Astron\'omico de Lisboa, \\
Tapada da Ajuda, 1349-018 Lisboa, Portugal\\
             \email{yun@oal.ul.pt, nuno@oal.ul.pt}
         \and
             Instituto de Ciencias del Espacio (CSIC) and Institut d'Estudis Espacials de Catalunya, \\ Facultat de F\'{\i}sica, Planta 7a, Universitat de Barcelona,\\ Avenida Diagonal 647, 08028 Barcelona, Spain \\
             \email{torrelles@ieec.fcr.es}
             }

   \date{Received 2007 February 7; accepted 2007 March 26}

 
  \abstract
  {}
   {Our new near-infrared $J$ and $K\!s$-band images taken towards IRAS~06468-0325 revealed the presence of extended emission with very unusual and enigmatic morphology: that of a diamond-ring, that is, a ring or torus with a bright source overlaping the edge of it (not inside it). We report the discovery of this source, explore its nature and propose possible interpretations of its morphological structure.}
   {We have observed IRAS~06468-0325 obtaining optical and infrared images through $IJHK\!s$ and $L'$ filters, $K$-band low-resolution spectroscopy, together with millimetre line observations of CO(1-0), $^{13}$CO(2-1), C$^{18}$O(2-1), and CS(2-1). }
   {Morphologically, IRAS~06468-0325 has two components: a bright, close to point-like source (the diamond) and a sharp-edge ring-like structure (the ring). The source is not detected in the optical, at wavelengths shorter than the $I$-band. The diamond is seen in all the imaging bands observed. The ring-like structure in IRAS~06468-0325 is clearly seen in the $I$, $J$, $H$, and $K\!s$. It is not detected in the $L'$-band image. Infrared colours of the diamond are compatible with excess circumstellar emission and a young stellar nature. A strongly non-gaussian and moderately bright CO(1-0) and $^{13}$CO(2-1) lines are seen towards IRAS~06468-0325, at v$_{LSR}$ of 30.5 km s$^{-1}$ (corresponding to a kinematic distance of 3 kpc). Very weak C$^{18}$O(2-1) and CS(2-1) lines were detected. $K$-band spectra of the diamond and of the ring are similar both in the slope of the continuum and in the presence of lines supporting the idea that the ring is reflected light from the diamond.}
   {With the current data, a few different scenarios are possible to explain the morphology of this object. However, the available data seem to favour that the morphology of IRAS~06468-0325 correspond to a young stellar multiple system in a transient stage where a binary co-exists with a circumbinary disc, similar to the case of GG Tau. In this case, the sharpness of the well-defined ring may be due to tidal truncation from dynamic interactions between components in a binary or multiple stellar system. IRAS~06468-0325 may be an important rare case that illustrates a short-lived stage of the process of binary or multiple star formation.}

   \keywords{ Stars: Formation --  Stars: pre-main sequence -- (Stars:) binaries: general -- (Stars:) circumstellar matter --  Stars: Individual (IRAS~06468-0325) -- (ISM:) reflection nebulae
               }

\titlerunning{ A ``diamond-ring'' star... }
\authorrunning{Yun, Torrelles, \& Santos}

   \maketitle
%

\section{Introduction}

The presence of nebular emission surrounding candidate young stellar objects (YSOs) detected in near-infrared images of high extincted molecular cloud cores has been taken as a clear sign of the young stellar nature of the objects (e.g. \cite{yun94, tapia97, rubio98}). 
The nebular emission seen is commonly associated to the phenomenon of mass ejection from young stars. This constitutes one of the major signposts of star formation in a molecular cloud occurring simultaneously with the accretion of material from the surroundings of the star.  The ejection of mass may be detected in the form of more or less collimated high-velocity winds, jets, and molecular outflows (e.g. \cite{mundt93}). The flow of gas ejected from the vicinity of the forming star results in the entrainment of ambient gas and the creation of cavities in the surrounding envelope and molecular cloud. 
The presence of these cavities allows radiation from embedded YSOs to escape via scattering off the walls of the cavities.
Thus, oftentimes the morphologies of the near-infrared nebulosities trace the walls of cavities excavated by the stellar jets and outflows (e.g. \cite{yun97, stark06}). In other cases, extended emission represents shocked gas at the location where the jets impact the ambient interstellar medium creating Herbig-Haro objects which are rich in (spectral) emission lines, seen in the optical (e.g. \cite{walawender05}), or in the near-infrared.
Finally, in some other cases, the presence of extended emission is due to binary (or multiple) star formation, with the presence of more or less extended nebulosity involving the multiple young stellar system.
In this case, the nebulae represent circumstellar and/or circumbinary structures (e.g. \cite{silber00}) whose evolution form multiple systems (e.g. \cite{bonnell94b}).

As part of an effort to produce a catalog of new YSOs identified in the IRAS data base, we have selected a sub-sample of candidate YSOs which were previously un-studied and obtained near-infrared images and millimetre line data towards these sources.
One of these objects (IRAS~06468-0325) is an odd-looking object. IRAS~06468-0325 is located in the outer Galaxy close to the Galactic plane in the direction of Monoceros.
Near-infrared $J$ and $K\!s$-band images taken towards IRAS~06468-0325 revealed the presence of extended emission.  But what makes this source rather striking is its very unusual and enigmatic morphology: that of a diamond-ring, that is, a ring or torus with a bright source overlaping the edge of it (not inside it). Furthermore, the ring is quite sharp and does not seem to follow any of the models of nebular emission from associated star formation regions or sources. 
In order to understand the nature of this source and what causes the morphological structures seen, we have performed an investigation of IRAS~06468-0325. 
We have collected additional data: low-resolution long slit $K$-band spectra, and $H$ and $L'$-band images.
We report in this article the results of this investigation.

In section 2 of this paper, we describe the large set of observational data collected so far in order to interpret the morphology of this source and shed some light on its nature. Section 3 presents the results. In section 4, we present a discussion of possible scenarios for the nature of this source. In Section 5, we summarize our findings.


\section{Observations and data reduction}

\subsection{Infrared imaging observations}

Near-infrared ($J$, $H$, and $K\!s$) observations were conducted during
2000 December 11 ($J$, $Ks$) and 2002 January 10 ($H$), using the ESO Antu (VLT Unit 1) telescope equipped with the short-wavelength arm (Hawaii Rockwell) of the ISAAC instrument. The ISAAC camera contains a 1024 $\times$ 1024 pixel near-infrared array and was used at a plate scale of 0.147 arcsec/pixel resulting in a field of view of 2.5~$\times$~2.5 arcmin$^2$ on the sky.  In the $J$ and $Ks$ bands, at each of 7 different jitter positions (with a jitter box of 30 arcsec), series of 10 images with 6~s exposure time were  taken. The total net integration time is therefore 7 min in the $J$ and $K\!s$ filters. In the $H$ band, at each of 7 different jitter positions (with a jitter box of 30 arcsec), series of 5 images with 6~s exposure time were taken. The total net integration time is therefore 3.5 min in the $H$ filter.

Broadband $L$ infrared images of IRAS~06468-0325 
were obtained on 2004 March 7, in service mode, with the long-wavelength arm 
(Aladdin) of the ISAAC instrument mounted on Unit 1 (Antu) of the ESO Very Large
 Telescope (VLT). The ISAAC camera (\cite{moorwood98}) contains a 1024 $\times$ 
1024 pixel and was used with the pixel scale of $0.07''$ per pixel resulting in 
a field of view of 1.2~$\times$~1.2 arcmin$^2$ on the sky. The $L$ band filter u
sed at ESO on the ISAAC detector is effectively $L'$ (3.78$\mu$m with 15\% width).

The observations were carried in chopping and nodding mode, using the ISAACLW\_img\_obs\_AutoChopNod observation template of the Aladdin arm of the ISAAC instrument (\cite{cuby05}). The chopping throw was $10''$ at a frequency of 0.43 Hz. The integration time per frame was 0.11 sec. Each on and off image
 consisted of nine co-added frames. Each exposure included 30 chop cycles. A series of nod cycles following a standard ABBA pattern with a random jitter between each cycle. Twenty-eight such exposures were taken.

\subsection{K-band spectroscopic observations}

K-band spectra towards IRAS~06468-0325 were obtained on 2004 March 17, in service mode, with ISAAC at the VLT using the SW arm in spectroscopic mode and the low-resolution grating. A slit width of 0.6$''$ was used, resulting in a spectral resolution $R=\lambda/\Delta \lambda \approx 750$.
The observations were carried using the ISAACLW\_spec\_obs\_AutoNodOnSlit observation template (\cite{cuby05}) and ocurred with a 1.0$''$ seeing.  

The spectra were reduced with the ISAAC pipeline package ECLIPSE. First master flat fields were created using the {\tt isaac\_spc\_flat} routine. Co-added spectrum images were then created using the {\tt isaac\_spc\_jitter} routine. Full details of the ISAAC reduction pipeline can be found in \cite{jung06}.
The images were then trimmed and extraction of the spectra was done using the usual IRAF routines inside the {\tt onedspec} and {\tt echelle} packages. In this procedure, the size of the apertures used for the spectral extraction was defined manually. This approach was essential in this case, given the requirement of extracting a faint spectrum, and the fact that automatic finding procedures within the IRAF {\tt apfind} and {\tt aptrace} routines did not manage to correctly follow the spectral dispersion. Once the spectra extracted, the wavelength calibration was done using the available argon spectrum taken during the same night. 
Telluric features were partly removed by applying the IRAF task 
{\tt telluric} to 
the spectra and using the spectrum of a B4 V star whose spectral features 
(H I absorption lines) had been previously removed (careful veriﬁcation of the 
reality of the features in the ﬁnal spectra was performed by comparison with 
the non-divided spectra). 
No further processing (e.g. flux calibration) were done since we were only interested in comparing the position and identification of lines and other features in two spectra taken at the same time, between {\it each other} (see below).
Further details of these observations (e.g. slit position in the sky) are given below.

\subsection{Optical observations}

An $I$-band image towards IRAS~06468-0325 was taken at the 3.9~m Anglo-Australian Telescope on the night of 2002 February 14 using WFI at a platescale of 0.2295 arcsec pix$^{-1}$. The integration time was 450 seconds.
$I$-band photometry obtained from this image was calibrated by comparison with stars appearing in the USNO-B Catalog and in the DENIS Point Source Catalog.

\subsection{Millimetre line observations}

Millimeter observations at 115 and at 98 GHz (the J=1-0 rotational transition of CO and the J=2-1 rotational transition of CS) and at about 220 GHz 
(the J=2-1 rotational transitions of $^{13}$CO and of C$^{18}$O) were carried out toward IRAS~06468-0325 using the 15m Swedish-ESO Submillimeter Telescope (SEST) of the European Southern Observatory (ESO) during 2002 April and December. The half-power beamwidths of the telescope were, respectively, 45, 52, and 23 arcsec at the above observing frequencies. The main-beam efficiencies were 0.70, 0.72, and 0.50, respectively. The backend included a 1000-channel acousto-optical spectrometer with 43 kHz resolution ($\sim$ 0.15 km s$^{-1}$ at 115 GHz).
Each source was observed using dual beam switching (chop throw of 11 arcmin) 
with integration times of 1 minute in CO and $^{13}$CO, and of 4 minutes in CS and C$^{18}$O toward each position. 
The pointing error was found to be better than 6 arcsec. 
We obtained a 5$\times$5 map of CO(1-0) spectra centered at the IRAS~06468-0325 position, taken in an approximately full-beam spaced grid. At the centre position several CO and $^{13}$CO spectra were obtained and co-added.

Spectral line intensities were calibrated and corrected for atmospheric losses using the standard chopper wheel method to obtain the antenna 
temperature $T^{\ast}_A$.
Baselines were fitted and removed using standard procedures of the Continuum and Line Analysis Single-dish Software (CLASS) package developed at
Observatoire de Grenoble and IRAM Institute. 

   \begin{figure}
   \centering
   \includegraphics[width=8cm]{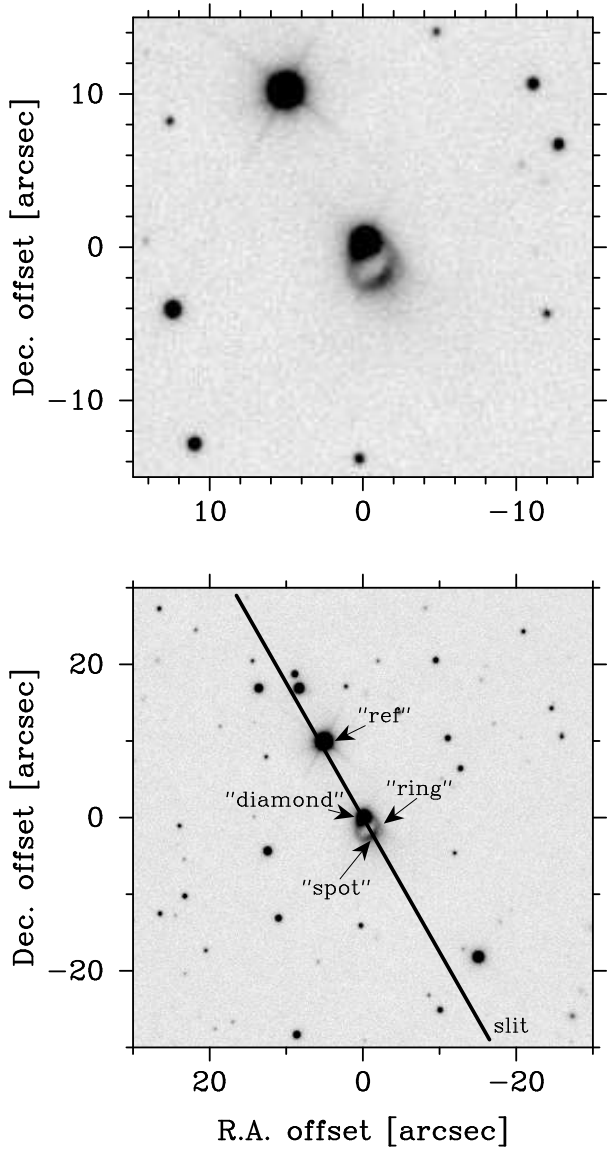}
   \caption{
{\it a) (top)} $K\!s$-band image towards IRAS~06468-0325 covering $30 \times 30 $ arcsec$^2$. Notice the ``diamond-ring" morphology of the source located at the center. 
{\it b) (bottom)} $K\!s$-band image towards IRAS~06468-0325 covering $1 \times 1 $ arcmin$^2$. The spatial location and orientation of the superimposed slit used for $K\!s$-band spectroscopy as well as the nomenclature used throughout this paper are indicated.
	} 
	\label{fig1}%
    \end{figure}
%

\section{Results}

Figure~1a shows the central region ($30'' \times 30''$) of the $K\!s$-band image obtained towards IRAS~06468-0325. Located at the centre of the image, the near-infrared counterpart of IRAS~06468-0325 has a very unusual morphology.
It immediately attracts attention as an extended source composed of a bright, relatively point-like component connected to a ring or torus of extended nebular emission with sharp edges. In other words, the near-infrared counterpart of IRAS~06468-0325 has the appearance of a diamond ring. We will designate the bright point-like component as the {\it ``diamond"} and the extended ring or torus as the {\it ``ring"} (see Figure~1b). The ring is not uniform in brightness. The brightest part appears as a region located at the other end of the diameter connecting the diamond and the center of the ring. We designate this brightest region of the ring as the {\it ``spot"}.

The diamond and the spot are separated by a projected angular distance of about $\sim 3''$ which is also the approximate angular diameter of the ring.
The diamond itself is not totally round and could be composed of a binary. However, the deviation from roundness is along the direction of the ring hampering the possibility of distinguishing whether it is caused by the presence of a close companion or by contamination from ring emission.
On the other hand, the spot does not correspond to the presence of a companion located at the brightest part of the ring. The spot is elongated following the ring and its FWHM ($2.90''$) is much larger than the typical FWHM ($0.44''$) of the point sources in the image.

   \begin{figure}
   \centering
   \includegraphics[width=8cm]{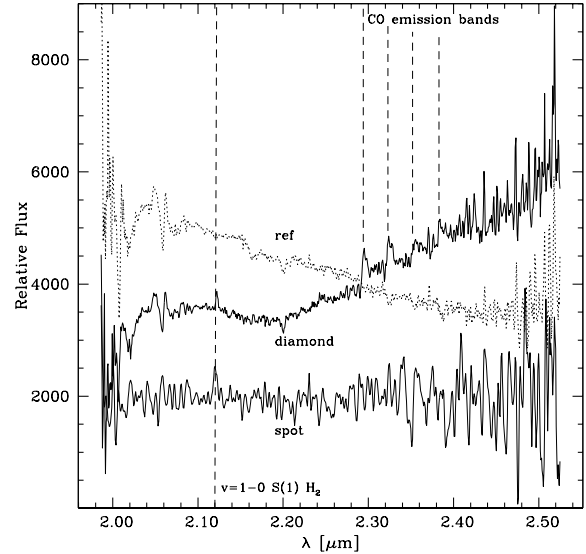}
   \caption{
$K\!s$-band spectra of the diamond (top solid line spectrum) and of the the ring spot (bottom solid line). The dotted-line spectrum is that of the reference field star. The spectrum of the ring spot has been multiplied by a factor of 30 for clarity purposes. The vertical axis gives fluxes relative to the telluric correction star.
The position of the $v = 1-0$ S(1) vibrational line of H$_2$ (at $\lambda=2.122$ $\mu$m) is marked by the dashed vertical line. The absorption features at around 2.06 $\mu$m are un-resolved telluric lines left-over from incomplete removal.
CO overtone bands (at $\lambda=$2.294, 2.323, 2.352, and 2.383 $\mu$m) are present in emission in the spectrum of the diamond.
	} 
	\label{fig2}%
    \end{figure}
%

The $K$-band spectra obtained are shown in Figure~2 where the spectrum of the spot has been multiplied by a factor of thirty for clarity purposes. The vertical axis gives fluxes relative to the telluric correction star. The location and orientation of the slit on the sky are shown on Figure~1b. They have been chosen to include both the diamond and the spot. By chance alignment, the slit included also the light of an additional bright field star (towards the northeast, marked as ``ref") which we have used as a reference field star. With a pixel scale along the slit of 0.147$''$, the three sources on the slit were well separated (20 pixels between the diamond and the spot; 77 pixels between the diamond and the ref star) in an average seeing of 1 arcsec.

We first compare the spectrum of the spot with that of the diamond. The two spectra are similar both in the slope of the continuum and in the presence of lines. In both spectra, a clear emission line is present, that of the $v = 1-0$ S(1) vibrational line of H$_2$  (molecular hydrogen line at $\lambda=2.122$ $\mu$m, marked in Figure~2 by the dashed vertical line). By contrast, the spectrum of the reference star can be seen to be very different and unrelated to the spectrum of the spot.
CO overtone bands (at $\lambda=$2.294, 2.323, 2.352, and 2.383  $\mu$m) are present, in emission, in the spectrum of the diamond. They may be present too in the spectrum of the spot, but their large widths and the higher noise level in this wavelength region prevents a firm conclusion.

  \begin{figure}
   \centering
   \includegraphics[width=8cm]{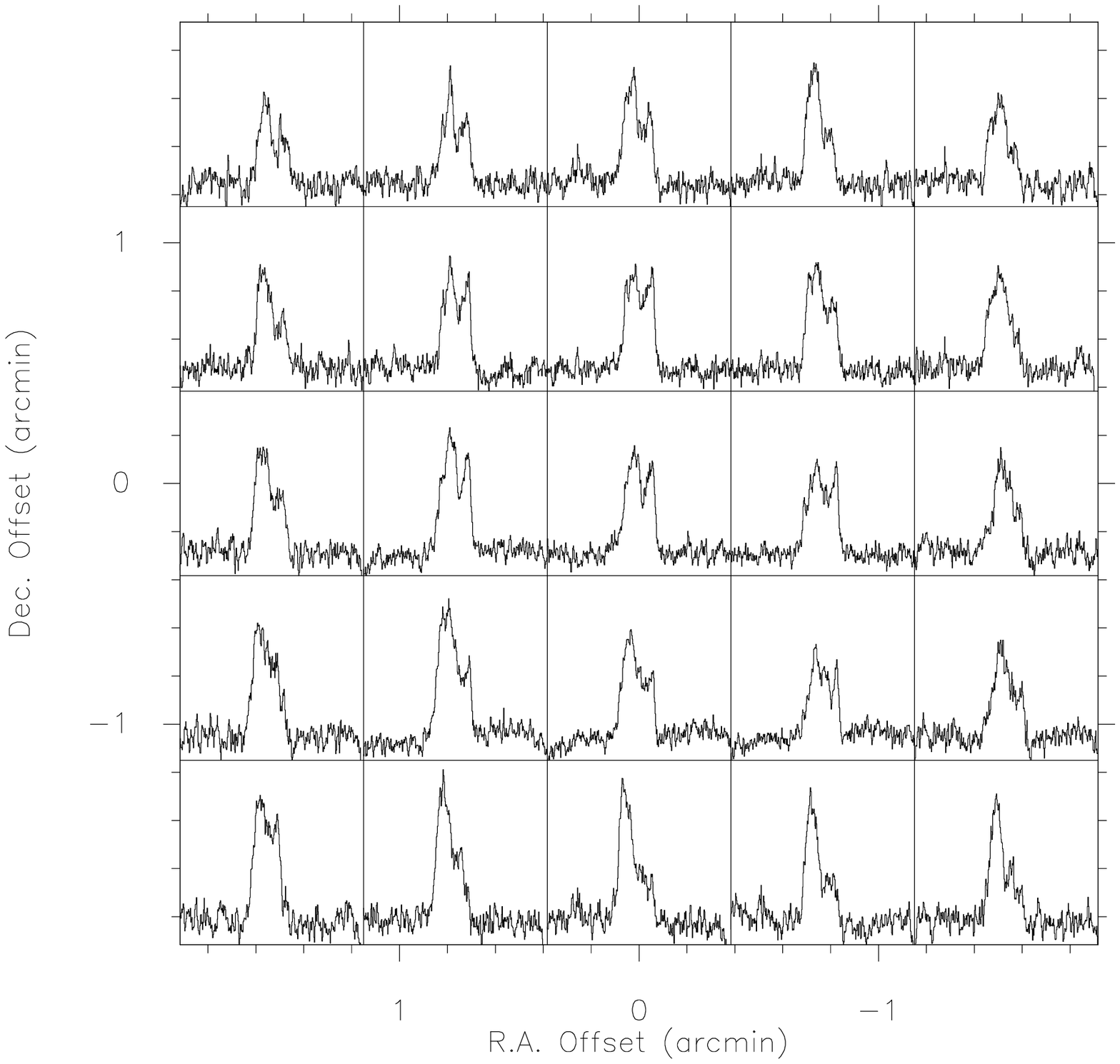}
	\vspace*{-4mm}
   \caption{Map of CO(1-0) spectra obtained towards IRAS~06468-0325. The spectra are presented in their relative angular positions.}
	\label{fig3}%
	\vspace*{4mm}
   \includegraphics[width=8cm]{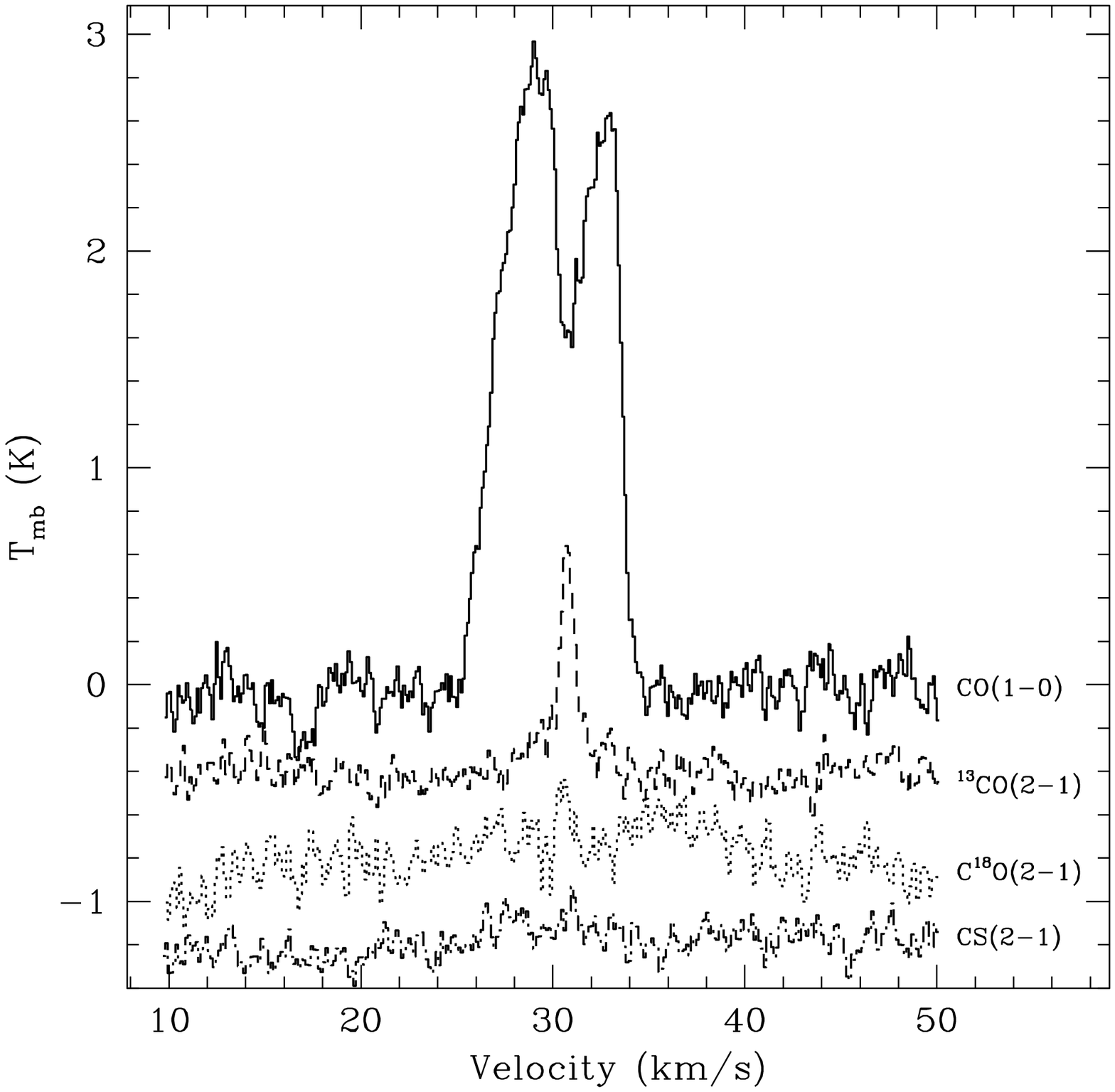}
	\vspace*{-4mm}
   \caption{
CO(1-0), $^{13}$CO(2-1), C$^{18}$O(2-1), and CS(2-1) spectra obtained towards 
IRAS~06468-0325 with SEST. The CO line (at LSR velocity of 30.5 km s$^{-1}$) appears non-gaussian. The $^{13}$CO(2-1) and C$^{18}$O(2-1)
lines were smoothed to the spectral resoltution of the CO(1-0) line.
}
	\label{fig4}%
    \end{figure}
%

The millimetre spectra are shown in Figures~3 and 4. The CO emission is resolved and a line has been detected towards all twenty-five mapped positions with similar intensities (Figure~3). The lines appear non-gaussian and double-peaked.
Figure~4 shows in more detail all the spectra obtained and co-added at the centre position. 
The CO(1-0) and $^{13}$CO(2-1) lines are moderately bright and appear at the LSR velocity of about 30.5 km s$^{-1}$ ($^{13}$CO peak and CO dip).
The C$^{18}$O and CS lines are very weak ($3\sigma$ detections) but they seem to peak at approximately the same velocity.

Figure 5\footnote{The $IJHK\!sL'$ images are available at the Centre de Donn\'ees Astronomiques de Strasbourg (CDS), at http://cdsweb.u-strasbg.fr.}  shows the $H$, $J$, $I$ and $L'$-band images towards IRAS~06468-0325.  The same ``diamond-ring" structure is seen in the $H$, $J$, and $I$-band images. The $H$ and the $J$-band images look very similar to the $K\!s$-band image. In the $I$-band image, both diamond and ring are seen but they are both much fainter when compared to the other sources in the field.
In the $L'$-band image, the ``diamond" is detected but not the ``ring". At the location of the ``spot" (the brightest region of the ring), a very faint ($2\sigma$) enhancement can be seen. 

Figure~6 (on-line) shows the $JHK\!s$ colour composite image towards IRAS~06468-0325. The red colour of the source is clearly revealed. The bluer colour of the spot, relative to the diamond, is also seen.

Aperture photometry of the ``diamond" and of the ``spot" components of IRAS~06468-0325 was carried out using standard IRAF tasks.
Photometry of the spot was performed integrating its surface brightness in the same aperture used for the photometry of the diamond.
 The results are given in Table~1. The $I$-band magnitude was estimated by comparison with stars appearing simultaneously in our image and in the USNO-B Catalog or the DENIS Point Source Catalog.

\begin{table*}
\caption{Photometry of IRAS~06468-0325}
\label{table:1}      
\centering                          
\begin{tabular}{c c c c c c c c c l}        
\hline\hline                 
R.A. & Dec & m${_I}$ & m${_J}$ & m${_H}$ & m${_K\!s}$  & m${_L}$ & $(J-H)$ & $(H-K\!s)$ & Obs.\\
(2000) & (2000) &  & & & & & & & \\
\hline                        
\noalign{\vspace{2 pt}}
06 49 23.1 & $-\,$03 28 54 &17.7& 15.7& 14.0 & 12.5 & 9.8 & 1.5 & 1.7 & diamond
\\
06 49 23.0 & $-\,$03 28 56 &17.7& 16.6& 15.9 & 15.5 & $>15.0$ & 0.7 & 0.4 & spot
\\
\hline\hline                                
\end{tabular}
\end{table*}


   \begin{figure}
   \centering
   \includegraphics[width=8cm]{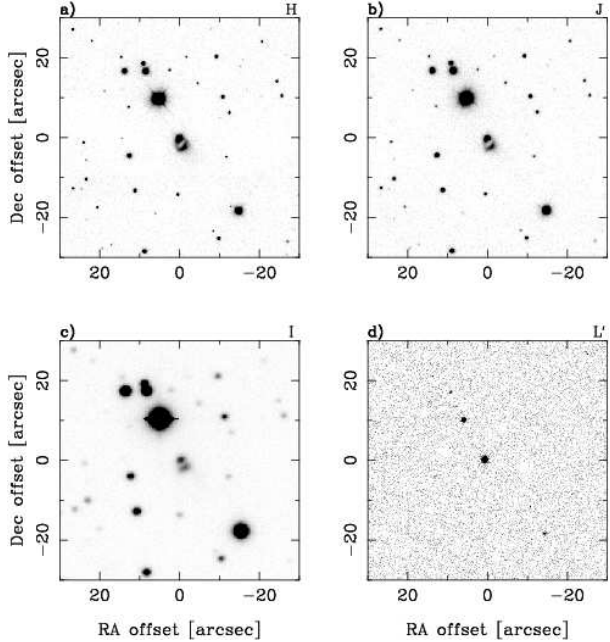}
   \caption{
$H$, $J$, $I$ and $L'$-band images towards IRAS~06468-0325 (located near the center) covering $1 \times 1 $ arcmin$^2$. The same ``diamond ring" morphology is seen in the $H$, $J$, and $I$-band images.}
	\label{fig5}%
    \end{figure}
%

   \begin{figure}
   \centering
   \includegraphics[width=8cm]{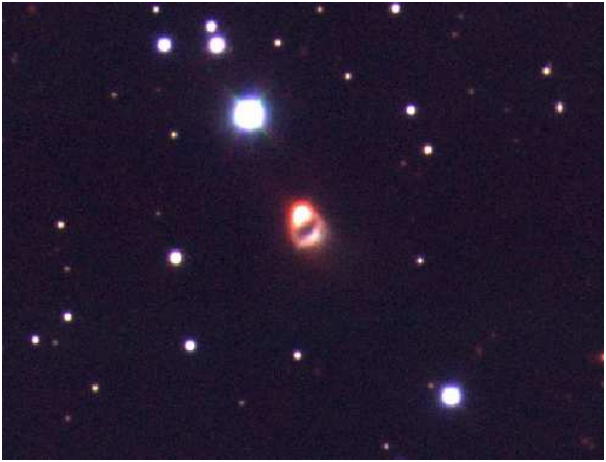}
   \caption{$JHK\!s$ colour composite image towards IRAS~06468-0325. The red colour of the source is clearly revealed.}
	\label{fig6}%
    \end{figure}


\section{Discussion: a diamond-ring star}

The set of data presented reveals that IRAS~06468-0325 appears as composed by a bright, close to point-like source (the {\it diamond}) and a ring or torus (the {\it ring}). 
This ring by itself has an outstanding morphology: that of an almost perfect circular or toroidal sharp-edged structure seen at an inclination angle.
The ring structure appears in all images ($I$, $J$, $H$, and $K\!s$-bands) except in the $L'$-band image (where however, there is a faint enhancement in brightness counts at the location of the {\it spot}). The lower sensitivity of the $L'$-band image together with the much bluer colour of the faint ring-like structure (see Table~1) may explain its non-detection in the $L'$-band. 
We point out that the $H$ and $I$-band images were taken on a different date from that of the $J$ and $K\!s$-band images, and that the $I$-band image was obtained with a different telescope and different instrument from the ones used to get the other images. This guarantees that no common structures seen on the images could be an artifact due to the instrument used.

This system is quite unusual and raises several issues which we will describe and address in the following sections.

\subsection{The molecular emission}

Is the molecular emission seen associated with the infrared source or does it just happen to be in the same line-of-sight? 
The data available are not conclusive. 
The CS and C$^{18}$O lines are weak but are likely to indicate the presence of some dense gas at this location which would associate them to the infrared source. 
In addition, the CO molecular map seems to exhibit a point-symmetry relative to the center position whose spectrum is the most symmetrical of the map. Higher signal-to-noise ratios are needed to better support this hint of point-symmetry in the current map.

The molecular emission displays Galactic velocities. An association of this emission to the infrared source supports the idea that the source seen in the $K\!s$-band image is likely to be a Galactic young embedded object.
The CO (1-0) lines (Figure~3) are mostly double peaked and display strongly non-gaussian character. This fact may argue in favour either of rotation or of the presence of turbulent gas motions (intermittent massive gas ejections? shocks?). However, there is no clear sign of a regular quiescent outflow (CO wings).
The set of spectra in Figure~4 shows that $^{13}$CO peaks at the velocity (30.5 km s$^{-1}$) of the CO dip. This suggests that the CO line profile is affected by self-absorption, possibly indicating a region of very high optical depth. 

Assuming a Galactic rotation curve that is essentially flat beyond the solar circle (\cite{clemens85}), and a Galactocentric distance to the Sun ($R_o$) of 8.5 kpc, we derive a kinematic distance of about 3 kpc to the source of the millimetre emission. At this distance, the radius of the ring corresponds to about 4500 A.U.

\subsection{The ``diamond": circumstellar environment and multiplicity}

The diamond appears as the reddest source in the image. Its near-infrared colours (Table~1) place it in a location on the $(H-K\!s)$, $(J-H)$ colour-colour diagram corresponding to sources with excess infrared circumstellar emission. The colours are also compatible with the source being in the Class~I stage of the evolutionary stages of low-mass YSOs introduced by \cite{adams87}. In this stage, sources are accreting matter from a circumstellar optically thick disc and are also surrounded by an infalling envelope. 
Given that the source is becoming optically visible (17th magnitude in the $I$-band), it is likely to be a late Class~I source, with the envelope in an advanced stage of dissipation. In fact, as has been shown by \cite{whitney03}, this classification depends on the inclination angle of the plane of the circumstellar disc relative to the line-of-sight.
Furthermore, the diamond is not totally round, showing an elongation. This could indicate the presence of a binary with the deviation from roundness being due to the presence of a close companion. 

Our $JHK\!sL$ photometry of the diamond gives a luminosity in these bands of $L_{\rm JHK\!sL}= 3.8 \, (\frac{D}{3000 \rm pc})^2 \, L_{\odot}$. If the entire IRAS flux comes from the diamond, its estimated bolometric luminosity is $L_{\rm bol}= 63 \, (\frac{D}{3000 \rm pc})^2 \, L_{\odot}$, where we have excluded the $IRAS 100$ $\mu$m flux (upper limit) from the luminosity calculation. The relatively low value of $L_{\rm bol}$ excludes the case of a massive star and suggests that this source be an intermediate or low-mass object.

The $K$-band spectrum of the diamond contains several emission features that are worthwhile noting. The emission features that can be un-ambiguously identified are the $v = 1-0$ S(1) vibrational line of H$_2$ (at $\lambda=2.122$ $\mu$m) and the CO overtones (at $\lambda=$2.294, 2.323, 2.352, and 2.383 $\mu$m). The presence of CO overtones in emission is not common in low-mass YSOs (\cite{greene96}). They are expected to be produced in a warm (2500-4500 K) and dense ($n \geq 10^{11}$ cm$^{-3}$) neutral gas. This gas could be located either in a circumstellar region such as the surface of the inner regions of optically thick circumstellar discs or in a high-velocity neutral wind (\cite{carr89}). Their excitation require special conditions for the temperature of the central star and for the disc accretion rate (\cite{calvet91}). For a low-mass star (with a low effective temperature and luminosity), CO emission may appear only for very low accretion rates.
For a given value of the accretion rate, the presence of CO emission bands imply that either the central star has a relatively high effective temperature (and luminosity), or another mechanism must be present for heating the disc or for producing CO emission elsewhere in the system. The presence of the molecular hydrogen line suggests the existence of a stellar wind which could heat the surface of the inner disc to excite the CO bands (\cite{greene96}). Additional data (e.g. higher spectral resolution and wavelength coverage) are needed to better precise the mechanism producing CO overtone emission in this source.

\subsection{The ring: reflected light}

Is the ring due to reflected light coming from the diamond? If so 
why is it so sharp and smooth ? 

The spectrum of the ring spot has the same features as the spectrum of the ``diamond" (see Figure~2) and therefore the enhanced brightness seen at this ring spot is likely to be due to reflection of light from the ``diamond" which  is illuminating a nearby structure that we see as a ring. 
This fact has an additional implication: it also rules out a chance coincidence of the lines-of-sight of the diamond and of the ring. These two components are undoubtly associated.

The flux ratio of the spot and the diamond increases with decreasing wavelength. The spot becomes increasingly brighter, relative to the diamond, as we go from $L'$ to $K\!s$, to $H$, to $J$, and to the $I$-band fluxes (see Table~1). Basically, this means that the spot is relatively blue, bluer than the diamond, as expected for scattered light. 
At the shortest wavelength observed ($I$-band), the brightness of the spot becomes similar to that of the diamond. Assuming that the spot is reflected light from the diamond (and therefore should not be equally bright), this implies higher extinction in the line-of-sight of the diamond than in the line-of-sight of the spot. This could be due to the presence of a circumstellar disc around the diamond which is also suggested by the infrared excess of the diamond (section 4.3). Interestingly, the inclination of the circumstellar disc that is required to produce extinction in our line-of-sight coincides (roughly, in qualitative terms) with the inclination that allows a good illumination of the ring spot: the circumstellar disc should be close to edge-on, obscuring the direction towards the observer but not the direction towards the ring spot.
In any case, the shape of the structure seen remains to be explained.

\subsubsection{The ring: a cavity?}

The envelope and the molecular cloud core where the source resides may have been excavated by a stellar jet and molecular outflow which are typically most intense during an earlier evolutionary stage (Class~0, \cite{andre93}) of the diamond. The presence of these cavities allows radiation from the central embedded object to escape via scattering off the walls of the cavities. Depending on the inclination angle of the cavity axis of symmetry to the line-of-sight, different shapes result for the nebulae that have been found associated to young embedded sources. This can be seen in models of embedded infrared nebulae that have been produced by \cite{lazareff90}, \cite{whitney93}, \cite{whitney03}, and Stark et al. (2006).

For an assumed distance of 3 kpc, the diameter of the ring is 9 000 AU much larger than the scale size of these models. However, given the uncertainty in the distance, we have attempted to match the morphology of the diamond-ring to the suite of models calculated by \cite{whitney03} and Stark et al. (2006) in order to estimate the physical properties of this source. 
Unlike common images of extended emission from other young stellar sources (e.g. \cite{yun95}, \cite{padgett99}), the diamond is not seen located within the boundaries of the extended emission but instead overlaps the ring itself (seen in projection). This excludes the models with low-inclination angles (seen close to face-on). A high inclination angle (together with the disappearance of one of the lobes of the models, say the redshifted lobe due to extinction effects) produces nebulae that are a bit more similar to the morphology of our source (see Figures~3 and 4 of \cite{stark06}).
However, none of the models seem to reproduce the ring structure. 
This could be due to the fixed shape of the cavities in the models (either curved or conical). The disc-only models (no envelope, no cavity) do not fit either.  Alternatively, this mismatch could be due to the fact that these models do not take into account the presence of stellar companions. We conjecture here that this source is not single but is likely to be composed of a multiple system (high angular resolution images are needed to try to spatially resolve this source and check for multiplicity).
Indeed, given the sharpness of the ring, truncation of a circumstellar/circumbinary structure (e.g. disc or torus) by the presence of companions appears as probable. That is, the morphology of this source could be explained by dynamic interactions between components in a binary or multiple stellar system. 

\subsubsection{The ring: a circumbinary structure similar to GG Tau?}

In fact, the most similar morphological structure that can be found in the literature is that of 
GG Tau. This is a well studied multiple (hierarchical quadruple or double binary) system (e.g.  \cite{roddier96}), with circumstellar discs (denounced by circumstellar excess infrared emission) and a circumbinary disc whose 90\% of the mass resides in a well-defined tidally truncated circumbinary ring  (\cite{guilloteau99}, \cite{silber00}).

In particular, the high angular resolution HST near-infrared image of GG Tau, taken by Silber et al. (2000), displays a point source and a sharp-edged ring and looks quite similar to IRAS~06468-0325. Compared to GG Tau, in order to reproduce the relative position of the diamond and the ring in our source, a higher inclination angle (closer to edge-on) seems necessary. A higher inclination angle should make the location of the central source appear closer to or overlaping a portion of the ring.

Thus, we may have found a larger version of GG Tau, which would confirm the presence of a system with several components (multiple system) with the sharpness of the ring being due to tidal truncation from dynamical interaction.
The dynamics of this multiple system might also have induced a motion of the diamond away from the centre of the ring. This shift may be needed, in addition to the larger inclination angle, to explain the location of the diamond apparently seen overlaping the ring.

Another object that resembles the diamond-ring structure is IRAS~04325+2402, a low-luminosity Class I (protostellar) source in the Taurus Molecular Cloud, described by \cite{hartmann99} (see their Figure~3 and object A/B in their Figure~7). Their A/B object includes a diamond-ring structure which represents just part of the nebulosity of the object. 
\cite{hartmann99} point out the similarity to GG Tau and mention the possibility of their ring being produced by the tidal action of an orbiting binary or multiple protostellar system. They also noted that their object A/B does not seem to lie inside the ring, but somewhat above it, just like our diamond-ring system. 

According to theoretical models of formation of binaries, the collapse of protostellar core material that initially forms the binary will leave a large fraction of the mass contained in the collapsing region. This remaining material, with high-angular momentum, will form a circumbinary disc or a tidal tail (\cite{bonnell94b}). Further evolution of this disc leads to its fragmentation and possible formation of a ring (\cite{bonnell94a}) and of additional stellar components (such as GG Tau, with a double binary). The system resembles the diamond-ring structure of IRAS~064680325 (if we degrade the infinite angular resolution of the model of Bonnell \& Bate (1994b) - see their Figure~2). 
Finally because this process is scale-free, it can occur for systems of any size, requiring only a relatively massive circumbinary disc, allowing for a wide system like our diamond-ring system.

The timescale for the evolution of a circumbinary disc is related to the orbital period of the binary. These short time-scales make it difficult to detect circumbinary discs observationally. Rare (short-lived) stages in the star formation process are hard to witness (e.g. FU Ori episodes, and effects of interacting binary forming stars).
Thus, IRAS~06468-0325 may be an important source in which to study circumbinary disc evolution.

\subsection{A post-main-sequence star?}

Given the rare morphological appearance of the source, other non young stellar objects cannot be ruled out from the observational data available. Thus, if it is a Galactic source, we could be in presence of a post-main-sequence source such as a planetary nebulae. In fact, these class of objects are known to produce exquisite nebulae with symmetrical shapes (e.g. \cite{manchado96}, \cite{balick98}, \cite{sahai98}, \cite{odell02}). However, again none of the models or images of planetary nebulae fit the shape of this source with the diamond overlaping the border of the ring. Furthermore, should this source be a post-main-sequence object, the relative brightnesses and colours of the diamond and the ring and the relatively large extinction (invisible in most of the optical wavelengths) would remain to be explained.

\section{Summary}

We have discovered an object with an unusual morphology in near-infrared images towards IRAS~06468-0325. It is composed of a bright, close to point-like source and a ring or torus.
This ring by itself has an outstanding morphology: that of an almost perfect circular or toroidal sharp-edged structure seen at an inclination angle.

This structure may represent scattered light from a cavity in the circumstellar envelope of the ``diamond". However, the ring is quite sharp and does not seem to follow any of the models of near-infrared nebular emission from single protostellar envelopes with cavities excavated by stellar winds, jets and outflows.

Alternatively, the morphology of IRAS~06468-0325 seems to be better explained  as a young stellar multiple system in a transient stage where a binary co-exists with a circumbinary disc, similar to the case of GG Tau. The sharpness of the well-defined ring may be due to tidal truncation from dynamic interactions between components in a binary or multiple stellar system. IRAS~06468-0325 may be an important rare case that illustrates a short-lived stage of the process of binary or multiple star formation.

\begin{acknowledgements}

We thank the fast and efficient action of the referee, as well as his/her constructive comments.
We thank the useful comments and discussions with Josep M. Paredes.
JLY thanks the staff of the Department of Astronomy and Meteorology of the University of Barcelona for hosting him during his sabbatical leave.
This work has been partly supported by the Portuguese Funda\c{c}\~ao
para a Ci\^encia e Tecnologia (FCT). JMT acknowledges partial financial support from the Spanish grant AYA2005-08523-C03.

\end{acknowledgements}

\end{document}